\documentclass[twocolumn,showpacs,preprintnumbers,superscriptaddress,amsmath,amssymb,showkeys]{revtex4-1}
\usepackage{graphicx}
\usepackage{dcolumn}
\usepackage{bm}
\graphicspath{{pict/}{}}

\newcommand\leqt[1]{\protect\label{eq:#1}}
\newcommand\reqtn[1]{\ref{eq:#1}}
\newcommand\reqt[1]{(\reqtn{#1})}

\begin{document}
\title{Coherent perfect absorption and reflection in slow-light waveguides}

\author{Nadav Gutman}
\email{nadav@physics.usyd.edu.au}
\author{C. Martijn de Sterke}
\affiliation{IPOS and CUDOS, School of Physics, University of Sydney, NSW 2006, Australia.}
\author{Andrey A. Sukhorukov}
\affiliation{Nonlinear Physics Centre, Research School of Physics and Engineering, Australian National University, Canberra, ACT 0200, Australia}
\author{Y. D. Chong}
\affiliation{Division of Physics and Applied Physics and Centre for Disruptive Photonic Technologies, Nanyang Technological University, Singapore 637371, Singapore}

\date{\today}

\begin{abstract}
We identify a family of unusual slow-light modes occurring in lossy multi-mode grating waveguides, for which either the forward or backward mode components, or both, become degenerate.  In the fully-degenerate case, by varying the wave amplitudes in a uniform input waveguide, one can modulate between coherent perfect absorption (zero reflection) and perfect reflection.  The perfectly-absorbed wave has anomalously short absorption length, scaling as
the inverse $1/3$ power of the absorptivity.
\end{abstract}

\maketitle

In periodically structured optical media such as Bragg gratings and photonic crystals, the propagation of light is strongly modified by coherent self-interference~\cite{Joannopoulos:2008:PhotonicCrystals}. This is the basis for numerous photonic devices, such as slow-light waveguides~\cite{Baba:2008-465:NPHOT}.  Slow-light waveguides are of considerable scientific and technological interest because light-matter interactions are enhanced in the slow-light regime; for applications involving light absorption and amplification, a short slow-light waveguide section integrated on a photonic chip can achieve the same functionality as a conventional bulk device~\cite{Soljacic:2002-2052:JOSB, Mork:2010-2834:OL, White:2012-43819:PRA, Grgic:2012-183903:PRL}. Slow-light waveguides containing loss and gain can also exhibit unidirectional and nonreciprocal light propagation~\cite{Poladian:1996-2963:PRE, Greenberg:2004-4013:OE, Lin:2011-213901:PRL, Feng:2013-108:NMAT}, with applications to photonic logic~\cite{Ginzburg:2009-4251:OE}. A key challenge in utilizing these waveguides is that the input coupling from non-slow-light waveguides tends to be very inefficient. This may be addressed by designing an impendence-matching transition region~\cite{Johnson:2002-66608:PRE, Vlasov:2006-50:OL, Velha:2007-6102:OE}, or by designing the slow-light waveguide itself to support evanescent modes which aid in impedance matching~\cite{White:2008-2644:OL, Spasenovic:2011-1170:OL}, but such techniques are typically formulated for the ideal case where the slow-light modes have real dispersion curves, without accounting for waveguide losses. Losses, however, can be considerable in slow-light modes; they may even be part of the intended functionality, as in the case of effective absorption arising from light-matter interaction effects.

In this Letter, we show that the problem of input coupling into slow-light waveguides can be substantially modified by absorption. We identify a family of complex slow-light modes in lossy multi-mode grating waveguides which can exhibit either coherent perfect absorption, or its opposite effect, perfect reflection.  Coherent perfect absorption is a phenomenon occurring in lossy cavities, in which illumination from all incident directions by an appropriately-chosen waveform leads to all the incident light being absorbed, with zero back-scattering. This concept was developed recently in a general optical scattering context \cite{Chong:2010-53901:PRL, chong2011hidden, Wan:2011-889:SCI, Noh:2012-186805:PRL}, as a generalization of critical coupling; interestingly, it has been previously discussed, in the context of lossy gratings, by Poladian~\cite{Poladian:1996-2963:PRE}. It has also been studied and demonstrated in thin metamaterial slabs by Zhang \textit{et al.}~\cite{Zhang:2012-e18:LSA}, who additionally pointed out that the ability to modulate the absorption from zero to unity by tuning the incident waveform has a range of practical applications.  Unlike these previous works, we consider a semi-infinite lossy waveguide, rather than a ``cavity'' like a finite waveguide segment.  Light is coupled into one end of the waveguide, and the interference of \textit{co-propagating} slow-light modes gives rise to perfect absorption or reflection.  Furthermore, coherent perfect absorption is associated with a degeneracy of the lossy waveguide modes, and in an appropriately-tuned waveguide this leads to anomalously short absorption lengths which scale as $\alpha^{-1/3}$, where $\alpha$ is the absorptivity of the waveguide medium, rather than $\alpha^{-1}$ like ordinary lossy modes.  In this way, a normally weakly-absorbing waveguide can absorb a correctly-chosen coherent input both perfectly (i.e., with zero reflection) and efficiently (i.e., with a short absorption length).

\begin{figure}
\centering\includegraphics*[width=\columnwidth]{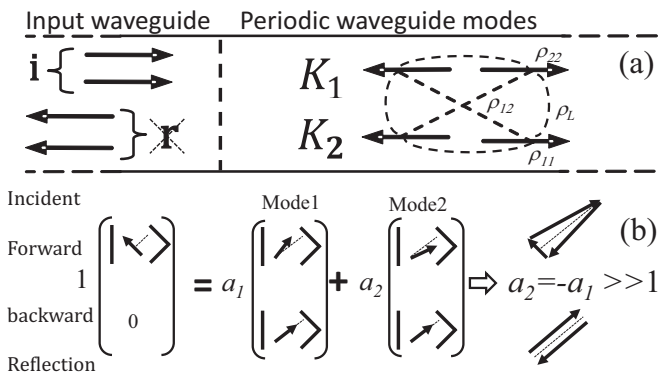}
\caption{(a) Schematic of a uniform waveguide coupled to a lossy periodic waveguide. (b) Schematic of the perfect absorption mechanism. The arrows represent the mode distribution of the forward and backward components on each side of the interface.  The input is normalized to 1, and the periodic waveguide has two excited modes with amplitudes $a_1$ and $a_2$.  }
\label{Fig1}
\end{figure}

Our structure consists of two joined semi-infinite waveguides, shown schematically in Fig.~\ref{Fig1}(a). On the left is a lossless uniform waveguide which supports two forward (input) modes and two backward (reflected) modes; these can be either spatial or polarization modes.  On the right is a lossy grating waveguide, which likewise supports two forward and two backward modes.  The forward grating modes are excited with amplitudes $a_1$ and $a_2$ respectively.  The backward grating modes have zero amplitude; in physical terms, this means that the grating waveguide's actual length greatly exceeds the lossy slow-light modes' absorption lengths, so that back-reflection from the far end is negligible.  The grating modes are combinations of the uniform waveguide's modes, which we refer to as ``basis modes''. Specifically, each grating mode consists of both forward \textit{and} backward basis modes.  Fig.~\ref{Fig1}(b) shows the principle by which reflection can be eliminated at the interface, resulting in coherent perfect absorption. The modes on each side are represented by vectors whose components correspond to forward- and backward-propagating basis modes. As indicated by the arrows, each forward or backward component is itself a vector in a two-dimensional modal subspace (since there are two basis modes for each direction of propagation). Now, suppose the forward grating modes have degenerate (linearly dependent) backward components.  For reasons which will become clear, we refer to this as {\sl partial mode degeneracy}. If the mode amplitudes satisfy $a_1 = -a_2$, the backward components destructively interfere; then the reflection vanishes and the incident light is perfectly coupled into the grating waveguide and absorbed.   As we will show, the amplitudes $a_{1,2}$ are much larger than the input amplitude, so the field intensity is strongly enhanced in the region of the lossy waveguide near the interface. Conversely, the forward components can be made degenerate, resulting in perfect reflection.  \textit{Full mode degeneracy}, i.e.~degeneracy of both forward and backwards components, can also be achieved.  This occurs when the waveguide supports {\sl complex degenerate bands} (CDB) satisfying
\begin{equation}\label{eq:CDB}
   (\omega-\omega_D)=(k-k_D)^{2}/\xi,
\end{equation}
where $\omega_D$ is a real frequency, and $k_D$ and $\xi$ are imaginary. Near the degeneracy point, the group velocity $v_g \equiv \partial\omega/\partial [\textrm{Re}(k)]\ll c$, so these are indeed slow light modes. When $\omega = \omega_D$, the group velocity vanishes, and we will see that these ``frozen light''~\cite{Figotin:2006-66613:PRE} modes exhibit strongly enhanced absorption.

The above argument can be developed using the coupled-mode theory of grating modes \cite{Sukhorukov:2007-17954:OE, Gutman:2012-33804:PRA}.  In the uniform waveguide, the modes have propagation constants $\tilde{K}_{1,2}$; there are forward ($+$) as well as backward ($-$) modes, and the total field is characterized by four mode amplitudes $E^{\pm}_{1,2}$. For convenience, we normalize the group velocities of these basis modes to $v=1$, though this is not essential~\cite{Sukhorukov:2007-17954:OE}. In the grating waveguide, losses induce $\exp(\pm\alpha_{1,2} z)$ decays for the basis modes; for convenience we take $\alpha_1=\alpha_2\equiv\alpha$. The grating also couples the basis modes, leading to amplitude variations which can be described by the envelopes $E_{1,2}^\pm(z,t)$. Assuming solutions of the form $\mathbf{E}\exp(i k z - i \omega t)$, where $\omega$ is the operating frequency, the wavenumbers $k$ are the eigenvalues of the ``wavenumber matrix''\cite{Sukhorukov:2007-17954:OE, Gutman:2012-33804:PRA}:
\begin{equation}\label{eq:Theta}
\mathbf{\Theta} = \left(
\begin{array}{cccc}
     \omega+i\alpha   &       \rho_{L}           &     \rho_{11}          &   	 \rho_{12}                   \\
       \rho_{L}      &  \omega-\delta+i\alpha    &     \rho_{12}          &    \rho_{22}                   \\
      -\rho_{11}      &      -\rho_{12}           &   -\omega-i\alpha     &    -\rho_{L}                   \\
      -\rho_{12}      &      -\rho_{22}           &     -\rho_{L}         &    -\omega+\delta-i\alpha
\end{array}
\right).
\end{equation}
The $\rho_{ij}$'s are integrals of the transverse mode profiles, and describe the coupling between basis mode~$i$ and the counter-propagating basis mode~$j$ \cite{Sukhorukov:2007-17954:OE}.  Such couplings can be induced either by a periodic modulation in a high index material or by multiple shallow Bragg gratings; in the latter case, each coefficient is defined by a grating with period $2\pi/\textrm{Re}(\tilde{K}_i+\tilde{K}_j)$.  By appropriate choice of grating structure, we take each $\rho_{ij}$ to be real-valued \cite{Sukhorukov:2007-17954:OE}; the triangle inequality then guarantees that $\det(\rho) \equiv \rho_{11}\rho_{22}-\rho_{12}^2 > 0$.  The parameter $\rho_L$ describes the coupling between the co-propagating modes, which can be generated by a long-period gratings with period $2\pi/\textrm{Re}(\tilde{K}_1-\tilde{K}_2)$. We find that such couplings are not relevant to the present study, and so henceforth set $\rho_L=0$. We have also included a small relative detuning $\delta$ from the Bragg condition.  The eigenvectors of $\mathbf{\Theta}$ are written as $\mathbf{E}_n=[E^+_1,E^+_2,E^-_1,E^-_2]^T_n$, where the $\pm$ superscripts denote forward and backward components. The eigenvalues $k_n(\omega)$ yield the waveguide's complex band structure.

\begin{figure}
\centering\includegraphics*[width=\columnwidth]{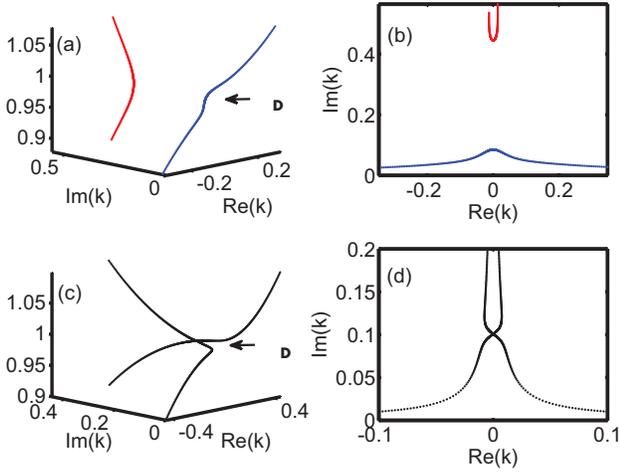}
\caption{
(Color online) Complex band structures of absorbing grating waveguides exhibiting coherent perfect absorption. (a, b): Partial mode degeneracy, with $\rho_{11}=1$, $\rho_{22}=0.9$, $\rho_L=0$, $\rho_{12}=0.2155$, $\delta=1.8937$ and $\alpha=10^{-2}$. (c, d): Full degeneracy (CDB), with $\rho_{11}=\rho_{22}=1$, $\rho_L=0$, $\rho_{12}=0.10016$, $\delta=1.99990$ and $\alpha=10^{-3}$.}
\label{Fig2}
\end{figure}

We now demonstrate how coherent perfect absorption arises in this system.  In the input waveguide, the amplitudes are $\mathbf{i}=[i_1,i_2]^T$ for the forward modes, and $\mathbf{r}=[r_1,r_2]^T$ for the backward modes [see Fig.~\ref{Fig1}(a)]. In the grating waveguide, the mode amplitudes are $a_{1,2}$ for the two forward grating modes, which have $\textrm{Im}(k) > 0$.  Matching these amplitudes gives
\begin{equation}\label{eq:Boundary}
    \textbf{E}_0
    =\left(
    \begin{array}{c}
        \mathbf{i}        \\
        \mathbf{r}
    \end{array}
    \right)
    =\left(
    \begin{array}{c}
        i_1        \\
        i_2        \\
        r_1        \\
        r_2
    \end{array}
    \right)
    =\mathcal{M}
    \left(
    \begin{array}{c}
        a_1        \\
        a_2        \\
        0          \\
        0
    \end{array}
    \right),
\end{equation}
where the columns of the matrix $\mathcal{M}$ consist of the eigenvectors of $\mathbf{\Theta}$. From $\mathcal{M}$, we can define a reflection matrix $\mathcal{R}$ which connects $\mathbf{i}$ and $\mathbf{r}$ by $\mathbf{r}=\mathcal{R}\mathbf{i}$:
\begin{eqnarray}\label{eq:R}
  \mathcal{R} =
  \begin{pmatrix}
        \mathcal{M}_{31}  & \mathcal{M}_{32}        \\
        \mathcal{M}_{41}  & \mathcal{M}_{42}
  \end{pmatrix}
  \begin{pmatrix}
        \mathcal{M}_{11}  & \mathcal{M}_{12}        \\
        \mathcal{M}_{21}  & \mathcal{M}_{22}
  \end{pmatrix}^{-1}
    = \mathcal{M}^-/\mathcal{M}^+.
\end{eqnarray}
Here, $\mathcal{M}^{\pm}$ are $2\times2$ matrices.  Zero reflection occurs when an eigenvalue of $\mathcal{R}$ is zero, i.e.:
\begin{eqnarray}\label{eq:detR}
    \det(\mathcal{R})=\det(\mathcal{M}^-)/\det(\mathcal{M}^+)=0.
\end{eqnarray}
The corresponding eigenvector of $\mathcal{R}$ yields the inputs $i_{1,2}$ which get perfectly absorbed.  Equivalently, $\det(\mathcal{M}^-)=0$, which occurs when the two columns of $\mathcal{M}^-$ are linearly dependent, i.e.~$\mathbf{\Theta}$ has two eigenvectors with equal backward components. Writing $\mathbf{\Theta}$ as
\begin{equation}
\mathbf{\Theta} = \left(
\begin{array}{cc}
     \mathcal{A} & \rho \\
    -\rho & -\mathcal{A}
\end{array}
\right),
\end{equation}
this partial degeneracy condition can be shown to be
\begin{align}\label{eq:PartialDegCon}
  \begin{aligned}
\det(&\mathcal{P})=0 \quad \textrm{where}\\
\mathcal{P}&= (\mathcal{A}+k_1 \mathcal{I}_2)^{-1}\, \rho \\
&\quad-(\mathcal{A}+k_2\mathcal{I}_2)^{-1}\,\rho\,(\mathcal{A}-k_2\mathcal{I}_2)^{-1}(\mathcal{A}-k_1\mathcal{I}_2).
  \end{aligned}
\end{align}
Here, $\mathcal{I}_2$ is the $2\times2$ identity matrix, and $k_1$ and $k_2$ are the wavenumbers of the two modes.  Plugging the explicit form for $\mathbf{\Theta}$ from Eq.~\reqt{Theta} into Eq.~(\ref{eq:PartialDegCon}), we obtain
\begin{equation} \leqt{PartialDegConA}
  (k_1 + k_2)^2 = -\, \frac{\rho_{12}^2\delta^2}{\det(\rho)}.
\end{equation}
Hence, the wavenumbers must satisfy $\rm{Re}(k_1)=-\rm{Re}(k_2)$ and $\rm{Im}(k_1) + \rm{Im}(k_2) = |\rho_{12}|\delta / \sqrt{\det(\rho)}$.

\begin{figure}
\centering\includegraphics*[width=\columnwidth]{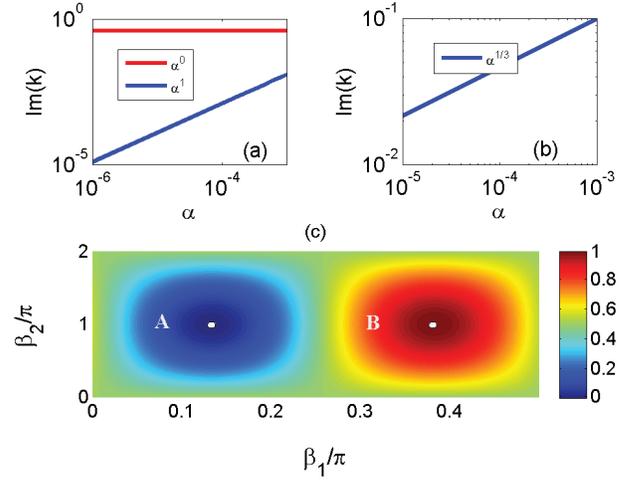}
\caption{
(Color online) Decay rate $\rm{Im}(k)$ vs.~absorptivity $\alpha$, for (a)~partial mode degeneracy and (b)~full degeneracy.   For each $\alpha$, the system is re-tuned by varying $\rho_{12}$ and $\delta$ so as to maintain partial mode degeneracy at some real $\omega_D$.  (c)~Reflection at full degeneracy vs. the relative amplitude ($\beta_1$) and phase ($\beta_2$) of the inputs, as parameterized in Eq.~(\ref{eq:i1i2}).
}
\label{Fig3}
\end{figure}

Fig.~\ref{Fig2}(a)-(b) show a complex band structure tuned to partial mode degeneracy at frequency $\omega_D$.  One mode has significantly smaller $\textrm{Im}(k)$ than the other.  At $\omega = \omega_D$, the two modes' values of $\textrm{Im}(k)$ approach one other, but do not meet.  The values of $\textrm{Im}(k)$ at the partial degeneracy point are plotted against the absorptivity $\alpha$ in Fig.~\ref{Fig3}(a).  At the interface $x = 0$, the two modes must have the same amplitude in order for the reflection to cancel exactly; hence, the intensity decay along the lossy waveguide is dominated by the less strongly-damped mode, whose value of $\textrm{Im}(k)$ scales linearly with $\alpha$.  (For the other mode, $\textrm{Im}(k)$ is larger and $\alpha$-independent.)

The absorption behavior is strongly modified if we tune the grating to achieve \textit{full} mode degeneracy.  In this case, the wavenumber matrix $\mathbf{\Theta}$ is defective; it cannot be diagonalized, only brought to Jordan normal form
\begin{equation}\label{eq:JordanBlockCDB}
\mathcal{J}=
\left(
\begin{array}{cccc}
k_D & 1   &      &     \\
    & k_D &      &     \\
    &     & -k_D &  1  \\
    &     &      &  -k_D
\end{array}
\right),
\end{equation}
where $\mathcal{J}=\mathcal{S}^{-1}\mathbf{\Theta}\mathcal{S}$ and $k_D$ is the complex wavenumber at the degeneracy.  Using the characteristic polynomial $p$, the condition for the Jordan normal form~(\ref{eq:JordanBlockCDB}) is
\begin{eqnarray}\label{eq:CDBcondition}
 p(k_D+\Delta k)&=&\det[\mathbf{\Theta}-(k_D+\Delta k) \mathcal{I}_4]  \nonumber \\
        &=&\Delta k^2(2k_D+\Delta k)^2 \;=\; \sum^4_{j=0}C_j\Delta k^j,
\end{eqnarray}
where $\Delta k=k-k_D$.  This requires the first two terms of the polynomial to vanish: $C_0=C_1=0$. We can use this to find the parameters for full mode degeneracy. It has previously been shown that mode degeneracies in lossless waveguides are associated with stationary points in real dispersion relations, of the form $(\omega-\omega_D)\propto(k-k_D)^{m}$ where $m$ is a positive integer~\cite{Gutman:2011-3257:OL, Figotin:2006-66613:PRE}. In the following, we will show that full mode degeneracy in a lossy structure yields the dispersion curves of Eq.~(\ref{eq:CDB}), featuring a \textit{complex} quadratic stationary point.  One such band structure is shown in Fig.~\ref{Fig2}(c)-(d).

According to Eq.~(\ref{eq:R}), fully-degenerate grating modes can be perfectly absorbed if $\det(\mathcal{M}^-)$ vanishes faster than $\det(\mathcal{M}^+)$ as we approach the degeneracy point.  Let us write the forward and backward components of the degenerate modes as $V_{k}^\pm = [E_1^\pm, E_2^\pm]^T$.  Expanding up to third order around the degeneracy, we obtain, from the perturbation theory of Jordan normal forms \cite{Lidskii:1966-73:RAR},
\begin{eqnarray}\label{eq:Vpm}
\mathbf{V}^{+}_{k_{1,2}}&=&
\mathbf{V}^+_0\pm\Delta k \mathbf{V}^+_1 +\Delta k^2 V^+_2\pm\Delta k^3 \mathbf{V}^+_3,\\
\mathbf{V}^{-}_{k_{1,2}}&=&\mathbf{V}^-_0\pm\Delta k\mathbf{V}^-_1+\Delta k^2 \mathbf{V}^-_2\pm\Delta k^3 \mathbf{V}^-_3.
\end{eqnarray}
The vectors are degenerate for $\Delta k=0$, and only the odd terms contribute to lifting the degeneracy. Perfect absorption requires $\mathbf{V}^-_1=0$. Hence, $\det(\mathcal{M}^-)\propto\Delta k^3$, since up to third order the vectors are identical. In contrast, for $\mathbf{V}^+_1\neq0$ we find $\det(\mathcal{M}^+)\propto\Delta k$. Thus,
$|\det(\mathcal{M}^+)|^2\propto\Delta k^2\propto\Delta \omega$ and $|\det(\mathcal{M}^-)|^2\propto\Delta k^6\propto\Delta \omega^3$, and
\begin{eqnarray}\label{eq:detRCPA}
    |\det(\mathcal{R})|^2&\propto&\Delta k^4\propto\Delta \omega^2.
\end{eqnarray}
Using Eq.~\reqt{Theta}, we find that full degeneracy is achieved when $\rho_{11}=\rho_{22}$ and $\delta = 2 \rho_{11} \rho_{12} (\alpha^2 + \rho_{12}^2)^{-1/2}$.
The degeneracy occurs at frequency $\omega_D = \delta/2$, with wavenumber
$k_D = [-(\alpha^4 + \rho_{12}^4 + \alpha^2 (\rho_{11}^2 + 2 \rho_{12}^2)) / (\alpha^2 + \rho_{12}^2) ]^{1/2}$.
In the limit of small absorption, $\alpha \rightarrow 0$, the degeneracy condition reduces to $\alpha \simeq \rho_{12}^3 / \rho_{11}^2$, with wavenumber $k_D \simeq i \rho_{12} \simeq \alpha^{1/3} \rho_{11}^{2/3}$. Hence, the absorption is increased compared to a homogeneous waveguide.  We have confirmed this scaling numerically, as shown in Fig.~\ref{Fig3}(b).

\begin{figure}[b]
\centering\includegraphics*[width=\columnwidth]{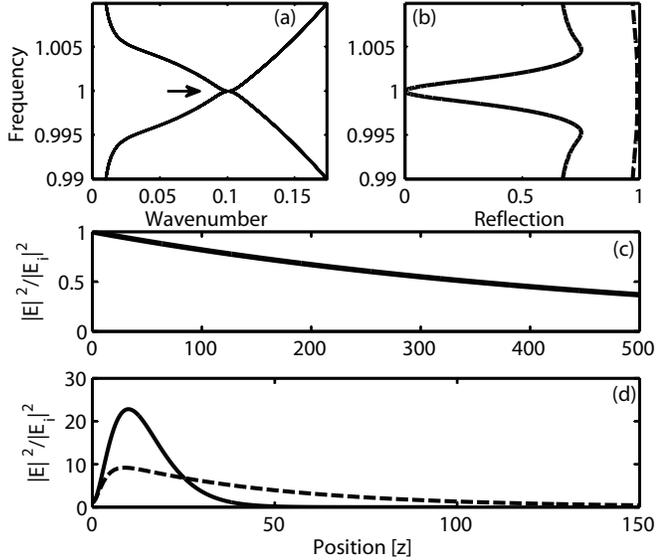}
\caption{
(a) Real band structure and (b) reflection spectrum near a full mode degeneracy point.  In (b), the two curves correspond to the input combinations which would yield perfect absorption (black) and perfect reflection (dashed) at the degeneracy, corresponding to the $\rm{A}$ and $\rm{B}$ points in Fig.~\ref{Fig3}(c). (c) Field intensity in a uniform absorbing waveguide. (d) Field intensity in a grating waveguide with partial mode degeneracy (dashed) and full degeneracy (black).  For both (c) and (d), the absorptivity is $\alpha=10^{-3}$.}
\label{Fig4}
\end{figure}

When the grating waveguide supports fully-degenerate modes, we can modulate between perfect absorption and perfect reflection using the input wave amplitudes. Let us parameterize the input amplitudes as
\begin{align}
    i_1=\cos{(\beta_1)}e^{i\beta_2}, \qquad i_2=\sin{(\beta_1).}
\label{eq:i1i2}
\end{align}
Fig.~\ref{Fig3}(c) shows the variation of the reflectance with $\beta_{1,2}$ at the degeneracy point.  At the marked point $\rm{A}$, the reflectance vanishes, whereas at point $\rm{B}$ it approaches unity.  The reflectance can be approximately written as $|r_1|^2+|r_2|^2=[1+\sin(2\beta_1)\sin(\beta_2+\gamma)]/2$, where $\gamma$ is a system-specific parameter; in this example, $\gamma=0.968\pi$.

The perfect absorption/reflection phenomenon can be explained by looking at the components of the degenerate modes. The incident fields $\mathbf{E}_0$ can be expressed as a linear combination of the modes' Taylor expansion: $\mathbf{i}_0=b_0 \mathbf{V}^+_0+b_1\mathbf{V}^+_-$. We find the amplitudes of the two degenerate modes $a_1,a_2\propto b_1\Delta k^{-1}$. When $\Delta k\rightarrow0$, the field in the lossy waveguide is
\begin{equation}\label{eq:Fields}
\mathbf{E}=b_1
\left(
\begin{array}{c}
 \mathbf{V}^+_1+i\mathbf{V}^+_0 z \\
 i\mathbf{V}^-_0 z
\end{array}
\right)
 e^{ik_Dz}.
\end{equation}
This grows polynomially before decaying exponentially with decay length $1/k_D$. At the degeneracy point, only $b_1$ (which represents the dependence of the input field on $\mathbf{V}^+_1$) is important. To maximize (minimize) the field, the input $\mathbf{i}$ should be parallel to $\mathbf{V}^+_1$ ($\mathbf{V}^+_0$); this corresponds to the points $\rm{A}$ (zero reflection) and $\rm{B}$ (total reflection) in Fig.~\ref{Fig3}(c), respectively.  The reflectances for the two input combinations are plotted in Fig.~\ref{Fig4}(b).  Fig.~\ref{Fig4}(a) shows the real part of the band structure, demonstrating that $\partial\omega/\partial k \rightarrow0$ at the degeneracy point.

Fig.~\ref{Fig4}(d) shows the field intensities inside a grating waveguide with partial and full mode degeneracy.  (For comparison, Fig.~\ref{Fig4}(c) shows the field intensity in a uniform absorbing waveguide.)  Both the partially and fully degenerate modes produce a slow-light intensity enhancement near the interface.  However, in the partially degenerate case the intensity subsequently decays at the same rate as in the uniform waveguide.  In the fully degenerate case, the decay length is significantly shortened.

In conclusion, we have shown that a multi-mode periodically-modulated lossy waveguide can be designed to feature a complex band degeneracy, which allows one to modulate between perfect absorption and perfect reflection by varying the input wave amplitudes.  In particular, the absorption can take place within a much shorter waveguide section than indicated by the underlying absorptivity, due to an anomalous scaling of the absorption length.  Although we have analyzed this phenomenon in the context of the coupled mode theory of grating waveguides, the basic physical principle is a general one, and it would be interesting to look for such effect in high-index periodic waveguides, such as photonic crystal waveguides suitable for on-chip slow-light applications.

This work was supported by the Australian Research Council programs (including Future Fellowship FT100100160 and Discovery Project DP130100086), by the Singapore National Research Foundation under grant No.~NRFF2012-02, and by the Singapore MOE Academic Research Fund Tier 3 grant MOE2011-T3-1-005.


%

\end{document}